\documentclass[structabstract]{aa}  
\usepackage{natbib}
\usepackage{graphicx}
\usepackage{lscape}
\usepackage{txfonts}
\begin{document}
   \title{The L\,1157 protostellar outflow imaged with the SMA}


   \author{A. I. G\'omez-Ruiz
          \inst{1,2}\fnmsep\thanks{Member of the International Max-Planck Research School for Astronomy and Astrophysics at the Universities of Bonn and Cologne. \email{arturogr@arcetri.astro.it}},
          N. Hirano\inst{3},
          S. Leurini\inst{2},
          S.-Y. Liu\inst{3}
          }

     \institute{INAF, Osservatorio Astrofisico di Arcetri, Largo E. Fermi 5, 50125 Firenze, Italy
         \and
   Max-Planck-Institut f\"ur Radioastronomie (MPIfR), Auf dem H\"ugel 69, 53121 Bonn, Germany
         \and
Academia Sinica,Institute of Astronomy \& Astrophysics, P.O. Box 23--141, Taipei, 106, Taiwan, R.O.C.
             }

   \date{Received xxxx xx, 2011; accepted xxxx xx, 2012}

 
   \abstract
   {The outflow driven by the class 0 low-mass protostar L1157-mm stands out for its peculiar chemical richness. However, its complex spatial/velocity structure makes difficult to interpret observations of different molecular tracers.}
   {The present work aims at mapping, at high spatial resolution, different molecular tracers that are important tools to study shocks and/or thermal-density structures in outflows.}
   {We use the Submillimeter Array to observe, at 1.4 mm, the blue-lobe of the L1157 outflow at high spatial resolution ($\sim$ 3'').}
   {We detected SiO, H$_2$CO, and CH$_3$OH lines from several molecular clumps that constitute the outflow. All three molecules were detected along the wall of the inner cavity that is supposedly related with the later ejection event. On the other hand, no emission was detected towards positions related to an old ejection episode, likely due to space filtering from the interferometer. The H$_2$CO and CH$_3$OH emission is detected only at velocities close to the systemic velocity. The spatial distributions of the H$_2$CO and CH$_3$OH are similar. These emission lines trace the U-shaped structure seen in the mid-infrared image. In contrast, the SiO emission is detected in wider velocity range with a peak at $\sim$14 km s$^{-1}$ blue-shifted from the systemic velocity. The SiO emission is brightest at the B1 position, which corresponds to the apex of the U-shaped structure. There are two compact SiO clumps along the faint arc-like feature to the east of the U-shaped structure. At the B1 position, there are two velocity components; one is a compact clump with a size of $\sim$1500 AU seen in the high-velocity and the other is an extended component with lower velocities. The kinematic structure at the B1 position is different from that expected in a single bow shock. It is likely that the high-velocity SiO clump at the B1 position is kinetically independent from the low-velocity gas. The line ratio between SiO (5--4) and SiO (2--1) suggests that the high velocity SiO clumps consist of high density gas of n$\sim$10$^5$ - 10$^6$ cm$^{-3}$, which is comparable to the density of the bullets in the extremely high velocity (EHV) jets. It is likely that the high-velocity SiO clumps in L1157 have the same origin as the EHV bullets.}
   {}

\titlerunning{L1157 imaged with the SMA}
\authorrunning{G\'omez-Ruiz, Hirano, Leurini, et al.}

   \maketitle
%

\section{Introduction}
There is now increasing evidence that extremely young outflows driven by class 0 low-mass protostars interact with surrounding ambient gas and produce strong shock waves. One prototypical and well-studied example of an outflow with strong shocks is the well-collimated bipolar outflow driven by the class 0 source of L$_{bol}\sim$ 11 $L_{\odot}$, IRAS 20386+6751, in the L1157 dark cloud at 440 pc from the Sun \citep[e.g.,][]{Ume92,Gueth97}. This outflow has been extensively studied with various molecular lines. The spatial-kinematic structure of the CO (1--0) was reproduced by a model of two limb-brightened cavities with slightly different axes \citep[]{Gueth96}. The locations of the two cavities are indicated by the yellow and green lines in Fig. 1. At the tips of these two cavities, labeled B2 and B1, respectively, strong SiO and NH$_3$ (3, 3) emission lines, which are considered to be good tracers of shocked molecular gas, are observed \citep{Gueth98,Zhang95,Zhang2000,Tafalla95}. On the basis of the observations of the highly excited ($J, K$) = (5, 5) and (6, 6) NH$_3$ emission lines, and the CO (6--5), (3--2), and (1--0) lines, the gas kinetic temperature in the shocked region at the B1 position was estimated to be $\sim$170 K, which is a factor of $\sim$10 higher than that of the quiescent gas \citep{Ume99,Hirano2001}. A remarkable correlation between the kinetic temperature and velocity dispersion of the CO $J$=3--2 emission along the lobe suggests that the molecular gas at the head of the bow-shock is indeed heated kinetically \citep{Hirano2001}.

In the shocked region, where the gas is significantly heated and compressed, various chemical reactions that cannot proceed in the cold and quiescent dark clouds are expected to be triggered off. \citet{Bachiller97} and \citet{Bachiller2001} surveyed molecular lines in the L1157 outflow, and found that molecules such as SiO, CH$_3$OH, H$_2$CO, HCN, CN, SO, and SO$_2$ are enhanced by at least an order of magnitude at the shocked region. In the blue lobe, molecular lines such as SiO, CH$_3$OH, and H$_2$CO mainly come from three regions labeled B2, B1, and B0 (Fig. 1). Furthermore, the spatial distributions of these shock-enhanced molecules differ from species to species; the SiO is remarkably enhanced at the B1 position, the CH$_3$OH is equally enhanced at both B1 and B2 positions, whereas HCO$^+$ and CN are enhanced in the region between the driving source and B1 position. This suggests that there is a stratification in the chemical composition of the shocked gas. 

Millimeter interferometric observations by \citet[]{Bene07} have resolved each emission structure (B0, B1, and B2) into a group of smaller scale clumps with a size of 0.02-0.04 pc. In addition, each emission structure also has its internal chemical stratification. At the B1 position, the HC$_3$N, HCN, CS, NH$_3$, and SiO lines are brighter in the eastern clumps, while the CH$_3$OH, OCS, and $^{34}$SO lines are brighter in the western clumps. Interferometric observations have also suggested a temperature stratification within the B1 position, with the highest temperatures found towards the apex of the B1 bow shock \citep{Codella09}.

 
Understanding the internal structure of the shocks in L1157 outflow is of particular interest in the light of the chemical studies currently underway with several facilities towards this remarkable object, for example Herschel key-program observations \citep{Codella10,Lefloch10}. In order to study the spatial and kinematic structure of the shocked gas at high angular resolution, we have mapped the blue lobe of the L1157 outflow at 1.4 mm using the Submillimeter Array (SMA). 


\section{OBSERVATIONS}

\subsection{SMA observations}
The observations were carried out on 2004 August 10 with the SMA\footnote{The Submillimeter Array is a joint project between the Smithsonian Astrophysical Observatory and the Academia Sinica Institute of Astronomy and Astrophysics, and is funded by the Smithsonian Institution and the Academia Sinica.} on Mauna Kea, Hawaii \citep{Ho04}. We used the compact-north array configuration that provided baselines ranging from 12.9\,m to 109.5\,m. The primary-beam size (HPBW) of the 6\,m diameter antennas at 217\,GHz was measured to be $\sim$54$''$. The entire region of the southern blue lobe was covered by four pointings separated by 30\arcsec (Fig. 1). The spectral correlator covers 2 GHz bandwidth in each of the two sidebands separated by 10 GHz. The frequency coverage was from 216.6 to 218.6 GHz in the lower sideband (LSB) and from 226.6 to 228.6 GHz in the upper sideband (USB). Each band is divided into 24 ^^ ^^ chunks'' of 104 MHz width. We used a uniform spectral resolution of 406.25 kHz across an entire 2 GHz band. The corresponding velocity resolution was 0.561 km s$^{-1}$. The visibility data were calibrated using the MIR software package, which was originally developed for Owens Valley Radio Observatory \citep{Sco93} and adapted for the SMA \footnote{http://cfa-www.harvard.edu/~cqi/mircook.html}. The absolute flux density scale was determined from observations of Uranus. A pair of nearby compact radio sources 1927+739 and 1806+698 were used to calibrate relative amplitude and phase. We used Uranus to calibrate the bandpass.


The calibrated visibility data were imaged using MIRIAD, followed by a nonlinear joint deconvolution using the CLEAN-based algorithm, MOSSDI \citep{Sau96}. We used natural weighting that provided a synthesized beam of 3\farcs4$\times$2\farcs3 with a position angle of 63$^{\circ}$. The rms noise level of the line data was 265 mJy beam$^{-1}$ at 1.3 km s$^{-1}$ spectral resolution. Four molecular lines, SiO(5--4), CH$_3$OH 4(2,2)--3(1,2)E, H$_2$CO 3(0,3)--2(0,2) and H$_2$CO 3(2,2)--2(2,1), were detected above a 4$\sigma$ level (see Table~\ref{freq}). All these lines were in the LSB, while no significant lines were detected in the USB. The continuum map was obtained by averaging the line-free chunks of the both sidebands. To improve the signal-to-noise ratio, the upper and lower sidebands were combined after the consistency of the images of two sidebands was confirmed. With natural weighting, the synthesized beam size was 3\farcs4$\times$ 2\farcs2 with a position angle of 63$^{\circ}$. The rms noise level of the 1.4 mm continuum map was 4.82 mJy beam$^{-1}$. 

\begin{table}
\caption{List of the detected transitions}
\begin{tabular}{lccc}
\hline
\hline
Transition & Rest Frequency & E$_u$ & vel. range\\
     & GHz & K & km s$^{-1}$\\
\hline
SiO (5--4)  &  217.10498 & 31.3 & $-$16.7 to $+$4.1\\
H$_2$CO 3(0,3)--2(0,2)  &  218.22218 & 21.0 & $-$3.7 to $+$4.1\\
H$_2$CO 3(2,2)--2(2,1)  &  218.47561 & 68.1 & $-$1.1 to $+$4.1\\
CH$_3$OH 4(2,2)--3(1,2)E &  218.44000 & 37.6 & $-$3.7 to $+$1.5\\
\hline
\label{freq}
\end{tabular}
\end{table}

\begin{figure}
\centering
\includegraphics[angle=0,width=9cm]{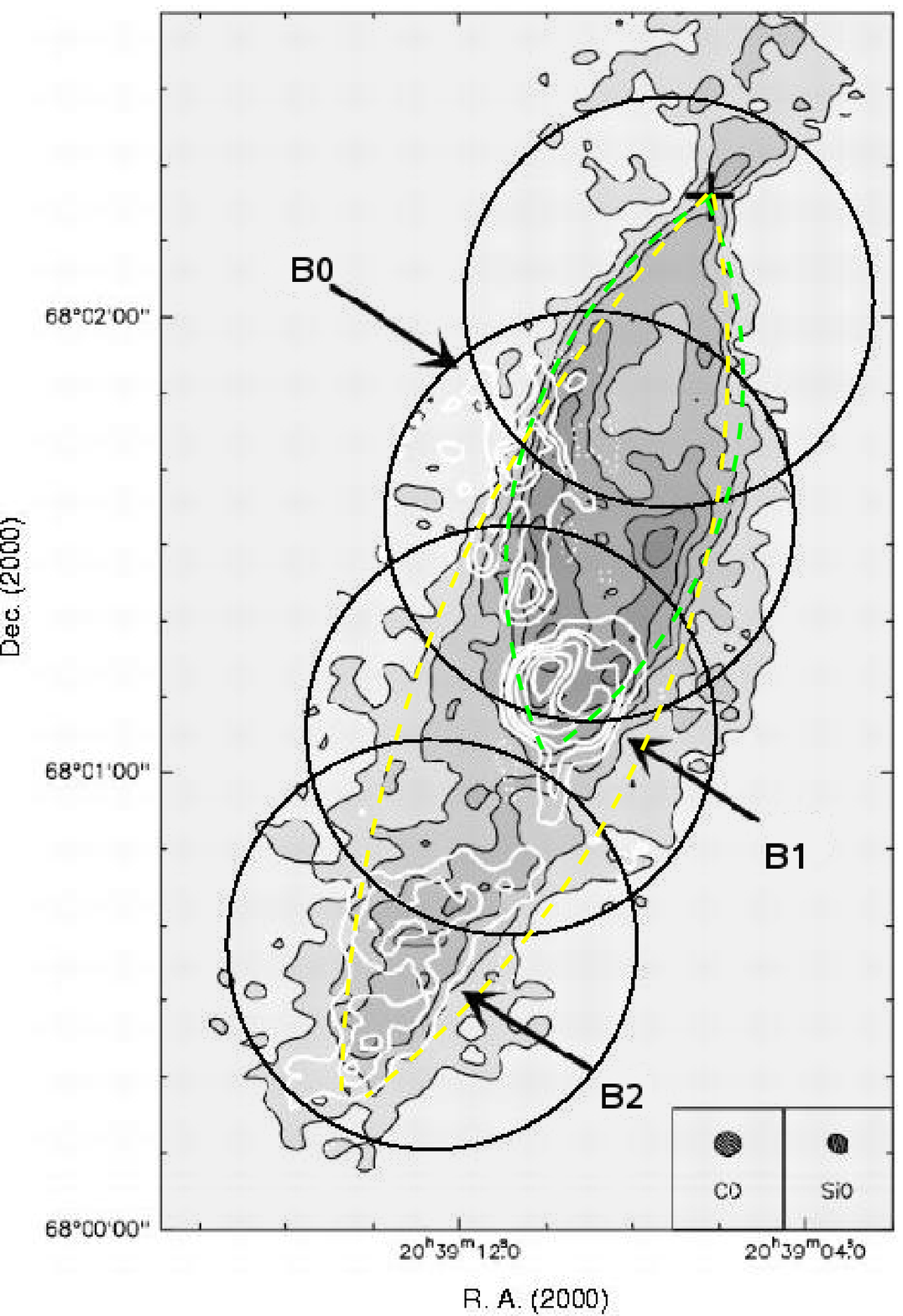}
\caption{The half-power primary beams of the 4 fields observed with the SMA (circles) superposed on the integrated CO (1--0) emission in grey scale and SiO (2--1) emission in the white contours \citep[from][]{Gueth98}. The cross indicates the position of the protostar L1157-mm, $\alpha = 20^h39^m06.19^s$, $\delta = 68^{\circ}02'15\farcs9$ (J2000.0) given by \citet{Gueth97}. The bottom-right corner shows the HPBW of \citet{Gueth98} observations. Yellow and green ellipses indicate the location of the two cavities proposed by \citet{Gueth97}.
\label{fig1}}
\end{figure}

\subsection{{\it Spitzer} IRAC observations}

Archival data of all four IRAC bands (3.6$\mu$m, 4.5$\mu$m, 5.8$\mu$m, 8.0$\mu$m) were retrieved from Spitzer data base
through Leopard. We have used the post-basic calibrated data (BCD) images for
our analysis. The mean FWHM of the point respond functions are 1\farcs66, 1\farcs72,
1\farcs88, and 1\farcs98 for bands 1, 2, 3 and 4, respectively. The details of
the IRAC observations of L1157 are described in \citet[]{Loon07}.

\begin{figure}
\centering
\includegraphics[bb=131 6 739 435,angle=0,width=9cm]{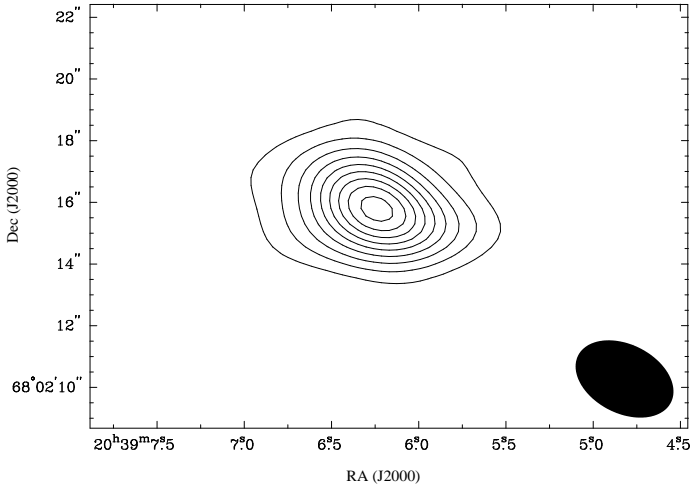}
\caption{The 1.4 mm continuum emission from L1157-mm. First contour and contour spacing is 3$\sigma$ ($\sigma$=4.82 mJy beam$^{-1}$), with the highest contour at 27$\sigma$. Black ellipse show the synthesized beam size (natural weighting). The parameters of a Gaussian fit to this image are presented in Table \ref{conti}.
\label{mmcont}}
\end{figure}

\begin{figure}
\centering
\includegraphics[angle=0,width=9cm,bb=141 90 355 438]{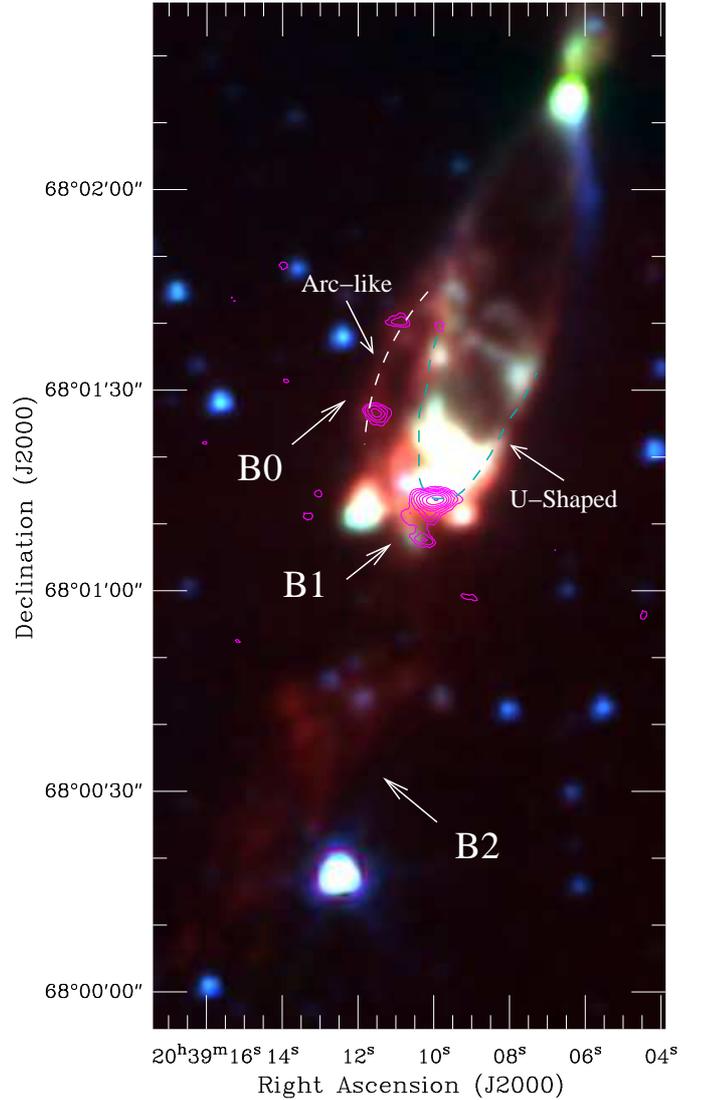}
\caption{Overlaid of SiO (5--4) integrated emission (integrated from $-$16.7 to $+$4.1 km s$^{-1}$; magenta contours) and IRAC 3 color image (blue: 3.6$\mu$m, green: 4.5$\mu$m, red: 8.0$\mu$m). White and cyan dashed lines indicates the arc-like feature and the U-shaped structure, respectively, delineated by the mid-IR emission. SiO contours start at 3$\sigma$ ($\sigma$=1.40 Jy beam$^{-1}$ km s$^{-1}$) and then separated by steps of 1$\sigma$.
\label{momtsio}}
\end{figure}

\section{RESULTS}

\subsection{Millimeter continuum emission}

The continuum emission map is shown in Fig. \ref{mmcont}. The continuum emission from the central source was detected at a level of $\sim$27$\sigma$. The parameters of the continuum source, obtained through an elliptical Gaussian fit, are listed in Table~\ref{conti}. The beam deconvolved size of the source is 2.7$"$ $\times$ 1.4$"$ (1200 x 600 AU) with a position angle of 88 deg. Previous observations at the same wavelength have been presented by \citet{Beltran04} and \citet{jorgensen07}. However, a comparison with these previous works is not always straightforward due to the different uv-coverage of the observations. The continuum map of \citet{Beltran04}, shows a spatially extended component with a size of $\sim$8$''$ around the compact component. On the other hand, the spatially extended component is not clearly seen in our map. This is probably because our map does not include the short-spacing data, which was added to the map of \citet{Beltran04}. 

\subsection{Mid-IR emission from the shocked gas}


A detailed analysis by \citet[]{Takami10} has shown that the mid-IR emission from this outflow is well explained by thermal H$_2$ emission excited by shocks.
Figure 2 shows a 3 color image of IRAC bands 1, 2 and 4 of the blue lobe of
the L1157 outflow, together with contours of the SiO (5--4) emission which will be described in the next section. The mid-IR emission delineates a
U-shaped structure with an apex near the position of B1. The eastern and western walls of the U-shaped structure are connected by an emission ridge. To the east of the U-shaped structure there is another fainter arc-like feature. The mid-IR emission from the B2 position is much fainter than that of the B1 position. 

The U-shaped structure in the mid-IR is confined inside the CO (1--0) cavity whose tip is B1. The blurred emission at B2 also has its counterpart in the CO (1--0) map. The location of the arc-like feature coincides with that of the B0 position. A comparison between the images of mid-IR and CO (1--0) implies that the arc-like feature corresponds to the eastern wall of the outer cavity produced by the B2 shock.


The difference in the Mid-IR color is considered to be due to the different excitation conditions \citep[]{Takami10}. The mid-IR emission at the B2 position and the arc-like feature is dominated by the longer wavelength component seen in the red color. This suggest that the excitation {of the H$_2$ molecule at} these two regions are rather low because of the lower temperature and/or density. Indeed this is confirmed by H$_2$ rotational line studies: the temperature derived from the H$_2$ line is lower than 300 K toward the arc-like feature, which is significantly lower than 1400 K at the tip of the U-shaped structure \citep[]{Nisini10}.


\begin{figure*}
\centering
\includegraphics[angle=0,width=16cm]{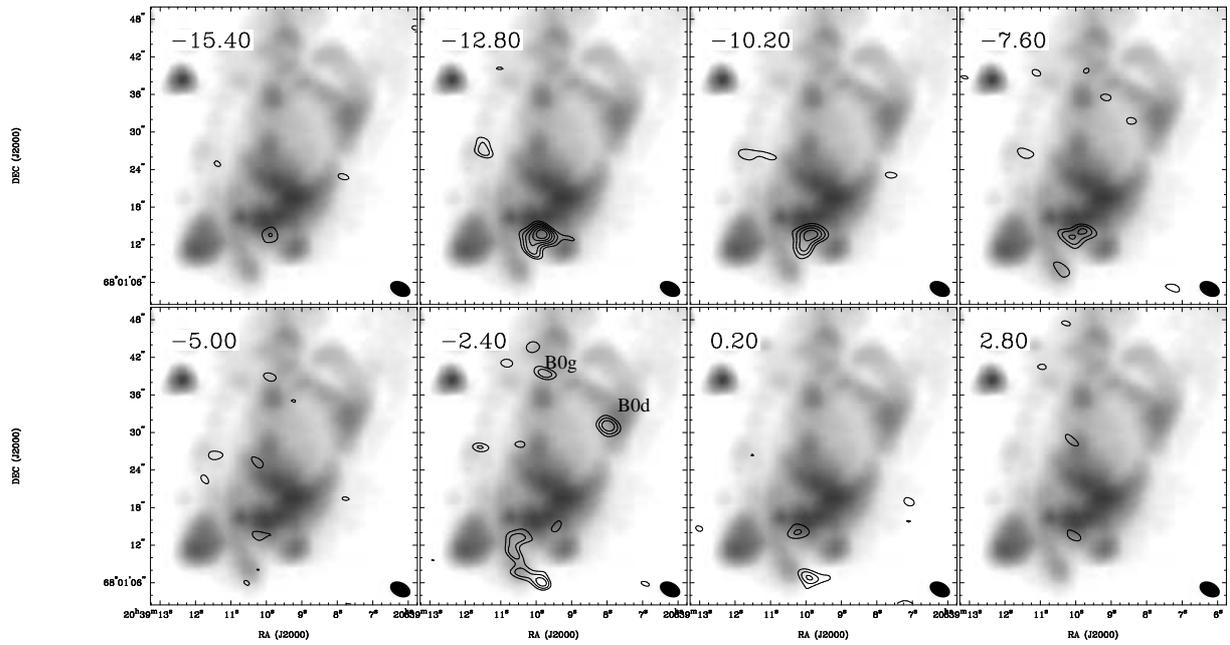}
\caption{The SiO (5--4) emission (contours) in channel maps of 2.6 km s$^{-1}$ width, overlaid on the 4.5$\mu$m emission from IRAC/Spitzer. SiO contours start at 3$\sigma$ ($\sigma$=0.188 Jy beam$^{-1}$) and then separated by steps of 1$\sigma$ up to a value of 8$\sigma$. The central velocity of each channel map is indicated in upper left, while the ellipse in the lower right shows the synthesized beam.
\label{siochans}}
\end{figure*}

\begin{figure*}
\centering
\includegraphics[angle=90,width=16cm]{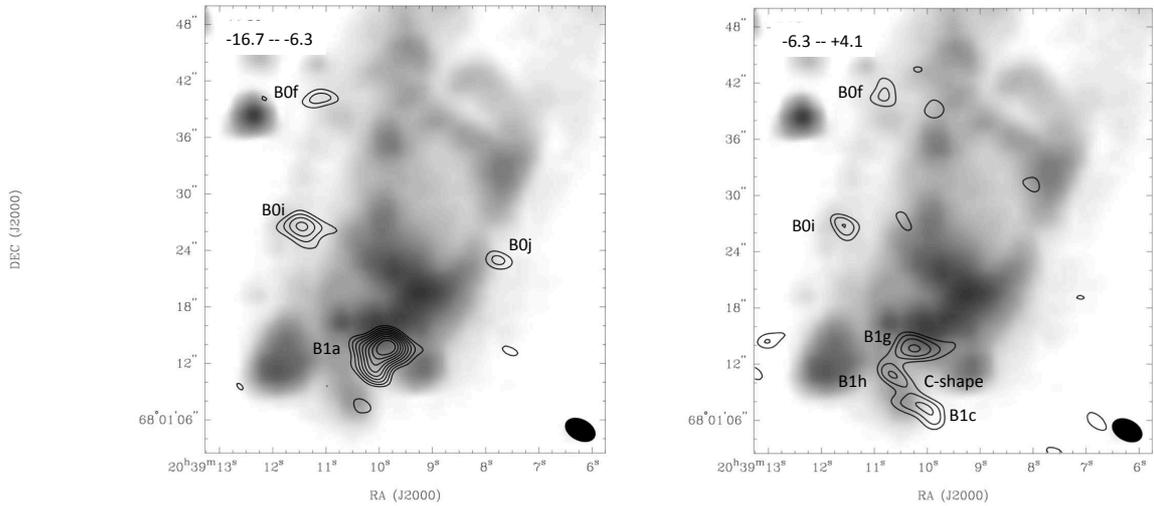}
\caption{The SiO(5--4) emission in the blue-lobe of L1157 outflow integrated in two velocity ranges. High velocity range (left panel) is integrated from -16.7 to -6.3 km s$^{-1}$, while low velocity range (right panel) from -6.3 to +4.1 km s$^{-1}$. The velocity range is indicated in the upper left corner of each panel. Contours start at 3$\sigma$ and then separated by steps of 1$\sigma$ ($\sigma$ = 0.985 Jy beam$^{-1}$ km s$^{-1}$). Grey scale represents the 4.5$\mu$m emission from IRAC/Spitzer.
\label{two-range-sio}}
\end{figure*}

\begin{table}
\caption{Parameters of the 1.4 mm continuum source}
\begin{tabular}{l c}
\hline
\hline
Peak intensity (Jy/beam) & 0.13$\pm$0.01 \\
Total integrated flux (Jy) & 0.21  \\
R.A. (J2000) & $20^h39^m06.24^s$ \\
DEC (J2000)     & $68^{\circ}02'15\farcs8$ \\
Deconvolved Major  axis (arcsec)  &    2.7 \\
Deconvolved Minor axis (arcsec)  &  1.4  \\
Deconvolved Position angle (degrees) &     88$\pm$4 \\
\hline
\end{tabular}
\label{conti}
\end{table}

\subsection{Molecular line emission observed with the SMA}

\subsubsection{SiO (5--4)}

The SiO 5--4 emission was detected in the velocity range from -16.7 km s$^{-1}$ to +4.1 km s$^{-1}$. Most of the SiO 5--4 emission observed with the SMA is blue-shifted with respect to the cloud systemic velocity (V$_{sys}$) of $V_{\rm LSR}{\sim}$ +2.7 km s$^{-1}$. The total integrated intensity map is shown in Fig. \ref{momtsio}, and velocity channel maps at 2.6 km s$^{-1}$ interval is presented in Fig. \ref{siochans}. The total integrated intensity map shows that the SiO (5--4) emission is brightest at the B1 position, which correspond to the apex of the U-shaped structure seen in the mid-IR. There are two compact SiO clumps along the faint arc-like feature to the east of the B0 position. Notably, no significant emission was detected at the position of B2, although the SiO (5--4) emission at this position was detected in the single-dish telescope observations of \citet[]{Bachiller2001}. Our observations with the shortest projected baseline of 9.3 k$\lambda$ were not able to recover the structure larger than $\sim$ 27$\arcsec$, while the emission component at the B2 position in the single-dish map is extended more than 20$\arcsec$. Furthermore, the r.m.s. noise level of our map is $\sim$4.6 K km s$^{-1}$, which is not sensitive enough to detect the SiO (5--4) emission at the B2 position, the peak intensity of which is $\sim$6 K km s$^{-1}$ in the single-dish map. Therefore this indicates that the SiO (5--4) emission from B2 position is spatially extended with low surface brightness and thus filtered out in our data.

\begin{figure}
\centering
\includegraphics[bb=133 12 699 552,angle=0,width=8cm]{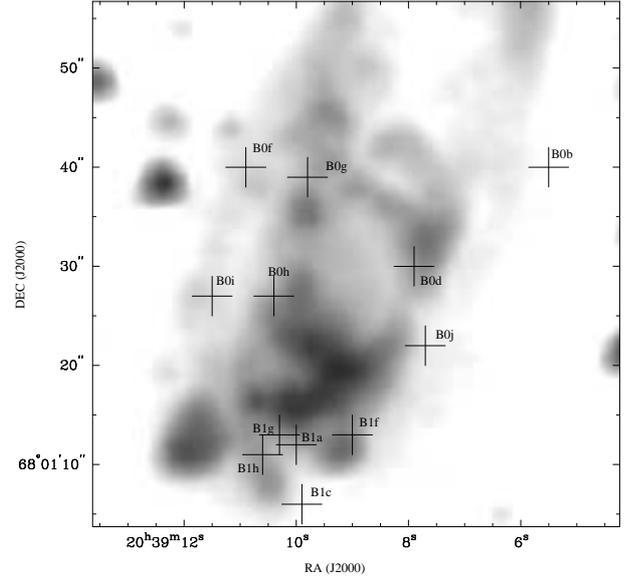}
\caption{A summary of all the clumps found in the SiO, H$_2$CO, and CH$_3$OH transitions reported in this paper. The crosses indicate the positions reported in Table \ref{clumps-tab}, overlaid on the 4.5 $\mu$m emission.
\label{all-clumps-pos}}
\end{figure}

The velocity-channel maps reveals that the most prominent clump at the B1 position appears in the velocity channels from -15.4 km s$^{-1}$ to -7.6 km s$^{-1}$ (the corresponding velocity offset from V$_{sys}$ is -18.1 km s$^{-1}$ to -10.3 km s$^{-1}$). On the other hand, the SiO (5--4) emission from the B1 position is more extended in the channels of -2.4 km s$^{-1}$ and 0.2 km s$^{-1}$ (the velocity offset is from -5.1 km s$^{-1}$ to -2.5 km s$^{-1}$). In Figure 5, we display the spatial distributions of the SiO in the velocity ranges from -16.7 km s$^{-1}$ to -6.3 km s$^{-1}$ and from -6.3 to +4.1 km s$^{-1}$. In these maps, all the SiO clumps detected above the 4-sigma level are labeled. The clumps B0d and B0g are visible only at the V$_{\rm LSR}$ of $-$2.4 km s$^{-1}$ in the channel maps. Therefore, these clumps are labeled in Fig. \ref{siochans}.  In Table \ref{clumps-tab} we list the clumps and their positions obtained by a two dimensional Gaussian fit. The list of clumps is shown in decreasing order of declination. We use the same notation as reported by \citet[]{Bene07} when our clumps are coincident within 4$\arcsec$ of their positions. Otherwise we name the clumps by following the sequence of letters for either B0 or B1. All the clump positions reported in Table \ref{clumps-tab} are summarized graphically in Fig. \ref{all-clumps-pos}. Three clumps shown in the map of the total integrated intensity (Fig. \ref{momtsio}) appear in the high velocity range (labeled B0f, B0i, and B1a). In the low velocity range, the most prominent feature is the C-shaped structure that consist of three clumps labeled B1c, B1g, and B1h. This C-shaped feature is surrounding the brightest clump B1a seen in the high velocity range. 

The SiO (5--4) emission from the B1 position was also observed by \citet{Gueth98} using the PdBI with a similar angular resolution (3.4$''$ $\times$ 2.9$''$). The overall features observed with the SMA at the B1 position are consistent with those of the PdBI map. Both SMA and PdBI maps show a compact clump in the velocity channels of $-$10.2 and $-$12.8 km s$^{-1}$ and the spatially extended feature in the lower velocity channels. In particular, the SMA and PdBI maps show good agreement in the peak flux of the compact component (B1a) at V$_{\rm LSR}$ = $-$12.8 km s$^{-1}$ channel ($\sim$1.5 Jy beam$^{-1}$), although B1a was located outside of the primary beam of the PdBI. In the low velocity channels, the SMA map does not show the linear structure ahead of B1c. This is probably because the surface brightness of this linear feature ($\sim$300 mJy beam$^{-1}$) is lower than our detection limit (3 $\sigma$ = 570 mJy beam$^{-1}$). The B1c clump in the PdBI map is more extended to the west as compared to the same component in the SMA map. Since the surface brightness of the extended part is above our detection limit, it is likely that the SMA could not detect the western part of B1c because of the limited u-v sampling.

\begin{figure}
\includegraphics[bb=34 15 725 405,angle=0,width=9cm]{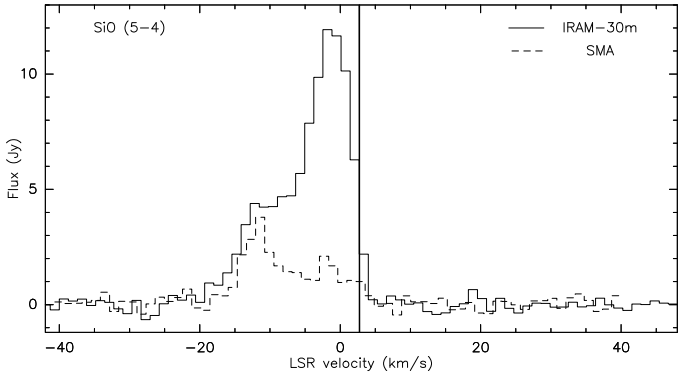}
\caption{SiO (5--4) spectra at the B1 position observed by IRAM 30m telescope (thin solid line) and the SMA (dashed line). Cloud velocity is indicated by a thick vertical line (V$_{\rm LSR}$=2.7 km s$^{-1}$). The SMA observations were convolved to the 11 arcsec resolution so as to match the beam of the IRAM 30m telescope. Spectral resolution is 1.3 km s$^{-1}$ in both cases.
\label{sio-spec-B1}}
\end{figure}

A number of observations have proven the particular chemical richness of the B1 position in L1157. Thus, the physical characterization at such position is important in order to account for kinematic effects in the chemical models that try to explain the emission from different molecular species \citep[e.g.][]{Viti11}. Therefore, most of the analysis of the SiO emission presented in the following sections are mainly focused on the B1 position. With the aim to estimate the missing flux at the B1 position, we have compared the single-dish spectrum taken with the IRAM-30m (provided by M. Tafalla) with our SMA spectrum convolved to the same angular resolution (11$\arcsec$). As shown in Fig. \ref{sio-spec-B1}, the single-dish spectrum in B1 peaks at V$_{\rm LSR}\sim$-2 km s$^{-1}$. On the other hand, the spectrum obtained with the SMA peaks at V$_{\rm LSR}\sim$-11 km s$^{-1}$, which is also seen as a bump in the single-dish spectrum. It is found that the SiO flux recovered by the SMA was only $\sim$10\% in the velocity range from -10.3 to +2.6 km s$^{-1}$, suggesting that most of the SiO emission in this velocity range comes from spatially extended region and filtered out in the current SMA map. In contrast, the SMA recovered up to $\sim$80\% of the SiO flux in the velocity range from -16.0 to -10.8 km s$^{-1}$. This implies that most of the SiO emission in this velocity range arises from the compact clump, B1a. Therefore, our SMA observations peaks at -11 km s$^{-1}$ instead of $\sim$-2 km s$^{-1}$, as the IRAM-30m observations, because most of the low-velocity emission is filtered out in the SMA map.

\begin{figure*}
\includegraphics[angle=0,width=18.5cm]{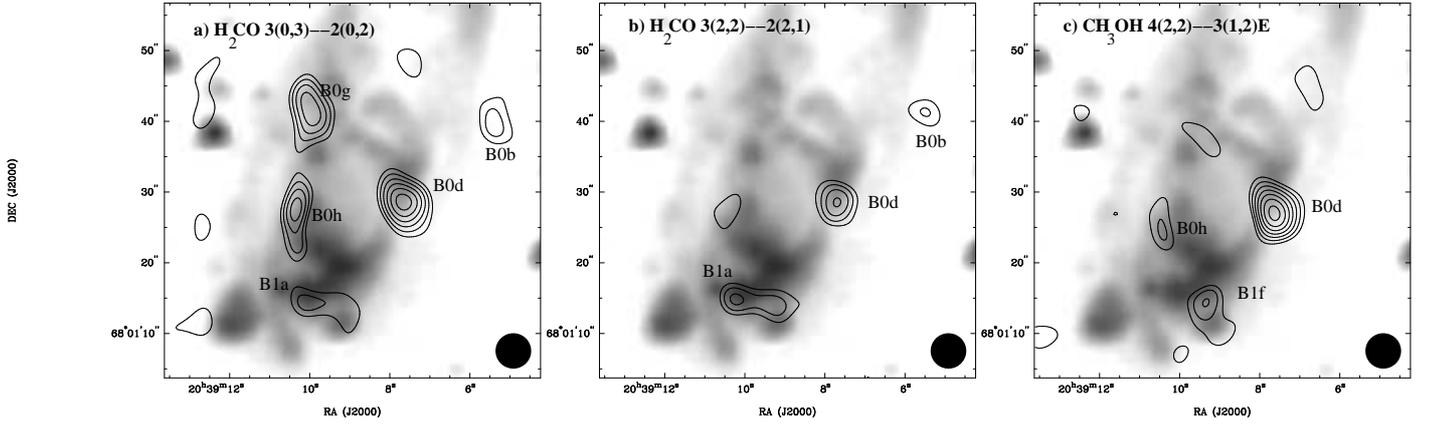}
\caption{Total integrated emission (see Table \ref{freq} for the velocity ranges) of the (a) H$_2$CO 3(0,3)--2(0,2), (b) H$_2$CO 3(2,2)--2(2,1), and (c) CH$_3$OH 4(2,2)--3(1,2)E transitions. In all panels, contours start at 3$\sigma$ ($\sigma$=1.35, 0.50, 0.55 Jy beam$^{-1}$ km s$^{-1}$, respectively) and then separated by steps of 1 $\sigma$. The grey scale represents the 4.5$\mu$m emission from {\it Spitzer}/IRAC. All maps are convolved to 5\arcsec$\times$5\arcsec (beam is shown at the bottom-right of each panel).
\label{two-range-h2co}}
\end{figure*}

\begin{table}
\caption{Clump positions}
\begin{tabular}{lccccc}
\hline
Clump & R.A. & Dec & \multicolumn{3}{c}{Molecular tracer\tablefootmark{a}}\\
      & J2000 & J2000 & SiO & H$_2$CO & CH$_3$OH\\
\hline
B0b\tablefootmark{*} & 20:39:05.5 & 68:01:40 & N & Y & N\\
B0f & 20:39:10.9 & 68:01:40 & Y & N & N\\
B0g & 20:39:09.8 & 68:01:39 & Y &  Y & N\\
B0d\tablefootmark{*} & 20:39:07.9 & 68:01:30 & Y & Y & Y\\
B0h & 20:39:10.4 & 68:01:27 &  N & Y & Y\\
B0i & 20:39:11.5 & 68:01:27 & Y & N & N\\
B0j & 20:39:07.7 & 68:01:22 & Y & N & N\\
B1g & 20:39:10.3 & 68:01:13 & Y & N & N\\
B1f\tablefootmark{+} & 20:39:09.0 & 68:01:13 & N &  N & Y\\
B1a\tablefootmark{*} & 20:39:10.0 & 68:01:12 & Y & Y & Y\\
B1h & 20:39:10.6 & 68:01:11 & Y & N & N\\ 
B1c\tablefootmark{*} & 20:39:09.9 & 68:01:06 & Y & N & N\\
\hline
\label{clumps-tab}
\end{tabular}
\tablefoottext{a}{Detections above 4~$\sigma$ marked Y, while non-detections marked N.}
\tablefoottext{*}{Reported also by \citet{Bene07}.} \tablefoottext{+}{Reported also by \citet{Codella09}.}
\end{table}

\subsubsection{H$_2$CO and CH$_3$OH}

Two formaldehyde (H$_2$CO) lines were detected in the LSB (see Table \ref{freq}). The H$_2$CO 3(0,3)--2(0,2) line was detected in the velocity range from -3.7 km s$^{-1}$ to +4.1 km s$^{-1}$, while the H$_2$CO 3(2,2)--2(2,1) line was detected from -1.1 to +4.1 km s$^{-1}$. In Fig. \ref{two-range-h2co}a,b we show the total integrated emission of both H$_2$CO transitions, overlaid on the IRAC 4.5$\mu$m emission. In order to improve the sensitivity, we have convolved both maps to 5$\arcsec$$\times$5$\arcsec$. As in the case of SiO, there is no significant H$_2$CO emission from B2 position. The spatial distribution of the H$_2$CO 3(0,3)--2(0,2) emission is similar to that of the H$_2$CO 3(2,2)--2(2,1), except the elongated structure along the eastern wall that is barely seen in the H$_2$CO 3(2,2)--2(2,1). The energy level in the upper state of the H$_2$CO 3(0,3)--2(0,2) transition is 21 K, while for the H$_2$CO 3(2,2)--2(2,1) is 68 K. Therefore, difference in the spatial distribution between the two transitions of H$_2$CO suggest that the excitation of the elongated structure in the eastern wall is lower than that of the B0d clump.

\begin{figure}
\includegraphics[angle=0,width=9cm,bb=130 15 700 551]{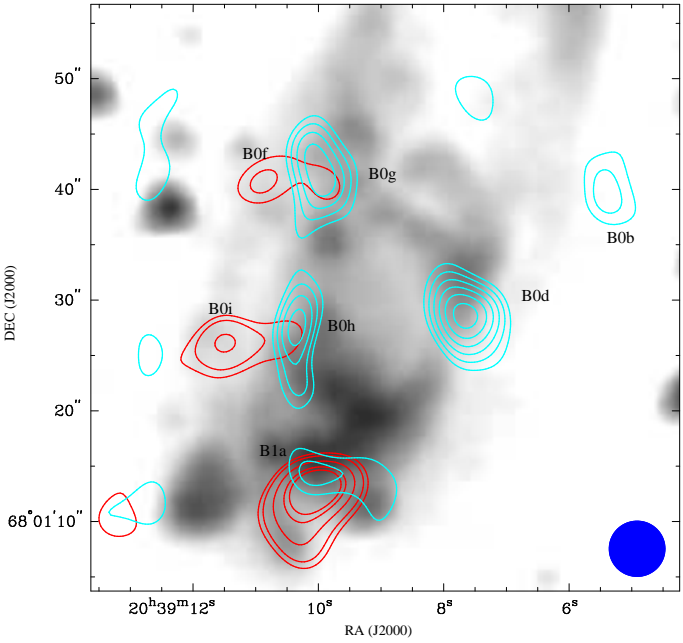}
\caption{Total integrated H$_2$CO 3(0,3)--2(0,2) emission (cyan; from -3.7 to +4.1 km s$^{-1}$) and the SiO (5--4) emission (red; from -16.7 to +4.1 km s$^{-1}$). Contours start at 3$\sigma$ ($\sigma$=1.35 and 2.60 Jy beam$^{-1}$ km s$^{-1}$, respectively) and then separated by steps of 1 $\sigma$. The maps are convolved to 5\arcsec$\times$5\arcsec (beam shown at the bottom-right). The grey scale represents the 4.5$\mu$m emission from {\it Spitzer}/IRAC. 
\label{3moles}}
\end{figure}

The H$_2$CO emission clumps detected above the 4$\sigma$ level are labeled in Figure \ref{two-range-h2co}, and their positions are listed in Table \ref{clumps-tab} (see also Fig. \ref{all-clumps-pos} for the distribution of all the clumps reported in Table \ref{clumps-tab}). We have also used the same notation as reported by \citet[]{Bene07} and \citet{Codella09} when our clumps are coincident within 4$\arcsec$ of their positions, otherwise define the new clumps in the same way as we have adopted for the SiO clumps.  As shown in Fig. \ref{two-range-h2co}a and \ref{3moles}, an elongated H$_2$CO 3(0,3)--2(0,2) emission feature, well traces the eastern wall (clumps B0g and B0h) of the mid-IR U-shaped structure. A Similar elongated feature was also seen in the HC$_3$N (11--10) map of \citet[]{Bene07} and the CH$_3$CN map of \citet{Codella09}. The brightest H$_2$CO emission component is seen in the western wall. This component is likely to be the counterpart of the B0d clump seen in the HCN (1--0) and CH$_3$OH (2$_K$--1$_K$) maps of \citet[]{Bene07}. The H$_2$CO clump at $\sim$ 15$\arcsec$ northwest of B0d is considered to be the counterpart of the B0b clump in the CH$_3$OH (2$_K$--1$_K$). 


The CH$_3$OH 4(2,2)--3(1,2)E line emission was found in the velocity range from $\sim$ -3.7 km s$^{-1}$ to +1.5 km s$^{-1}$. The CH$_3$OH total integrated intensity map shown in Figure \ref{two-range-h2co}c was also convolved to the 5$\arcsec$$\times$5$\arcsec$ resolution in order to improve the signal-to-noise ratio. The overall distribution of the CH$_3$OH 4(2,2)--3(1,2)E emission is similar to that of the H$_2$CO 3(2,2)--2(2,1). The CH$_3$OH emission is brightest in the B0d clump, but also showing emission at B0h and B1f. No CH$_3$OH emission was detected at the B2 position. The clumps detected above the 4$\sigma$ level in the CH$_3$OH 4(2,2)--3(1,2)E line are also listed in Table \ref{clumps-tab} (see also Fig. \ref{all-clumps-pos} for the distribution of all the clumps reported in Table \ref{clumps-tab}).


\section{DISCUSSION}

\subsection{Physical conditions of the clumps}


\subsubsection{The SiO high velocity clumps}


\begin{figure}
\centering
\includegraphics[angle=0,width=8cm]{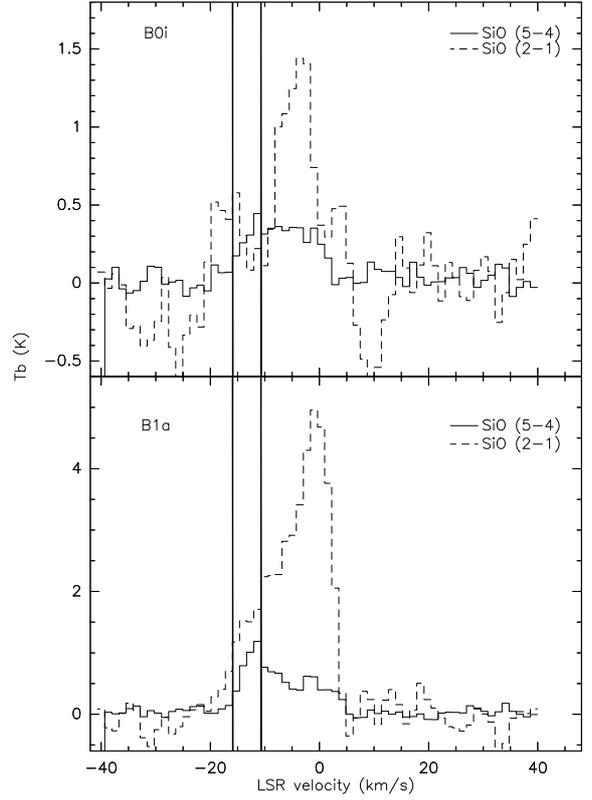}
\caption{SiO (2--1) and (5--4) spectra at B0i and B1a positions in brightness temperature scale. SiO (5--4) has been convolved to the angular resolution of the SiO (2--1) observations (i.e. 9.5$\times$8.0 arcsec). Thick vertical line indicates the high-velocity ranges on which the analysis has been done, from -16.0 to -10.8 km s$^{-1}$. 
\label{spect-HV-21-54}}
\end{figure}

Using the line ratio between the SiO (5--4) and the SiO (2--1) observed by \citet[]{Zhang2000} with a synthesized beam of 9\farcs5$\times$8\farcs0, we have derived the physical conditions of the gas in the high-velocity clumps. For our analysis we have selected the portion of the high-velocity range which is less affected by the missing flux. The analysis has been done at the B0i and B1a positions, on which the SiO (5--4) emission was detected above 5$\sigma$ level. We note that an analysis using several SiO lines observed with a single-dish telescope has been done by \citet[]{Nisini07}. However, the analysis of \citet{Nisini07} used single-pointing observations, which means that the data were taken with different beam sizes, e.g. $\sim$27\arcsec for SiO (2--1) and $\sim$11\arcsec for SiO (5--4). In addition, the line ratios used the total integrated intensity, including the low-velocity component. On the other hand, our analysis used higher resolution data, and focused on the high-velocity component.

Since the interferometric SiO (2--1) observations agreed to better than 20\% with the single-dish results, \citet[]{Zhang2000} concluded that most of the SiO (2--1) flux was recovered with the interferometer. Therefore we do not take into account the effect of the missing flux in this transition. Our SiO (5--4) observations were convolved to the angular resolution of the SiO (2--1) data (i.e. 9\farcs5$\times$8\farcs0). In Fig. \ref{spect-HV-21-54} we show the SiO (5--4) and (2--1) spectra at the position of clumps B0i and B1a in brightness temperature scale. The portion of the high-velocity range on which most of the single-dish SiO (5--4) emission at the B1a position is recovered by our SMA observations is -16.0 to -10.8 km s$^{-1}$. In this analysis, the velocity range for the B0i clump was assumed to be the same as that of the B1a clump. The linewidth was assumed to be 5 km s$^{-1}$ for both B1a and B0i. Since smoothing the SMA maps to $\sim$ 9$''$ beam averaged the emission of adjacent clumps, the spectrum at B1a includes the emission from B1g, B1h, B1c, and B1f. However, the emission from adjacent clumps appears only in the low-velocity range, and does not affect the analysis. The integrated line intensities of SiO (2--1) and (5--4) in this velocity range are 1.5$\pm$0.6 and 1.5$\pm$0.1 K km s$^{-1}$, respectively, at B0i, and 7.7$\pm$0.6 and 4.4$\pm$0.1 K km s$^{-1}$, respectively, at B1a. The derived (2--1)/(5--4) ratios are therefore 1.0$\pm$0.4 for B0i and 1.7$\pm$0.1 for B1a.





We use the non-LTE program RADEX \citep[]{Tak07} in the LVG approximation and plane parallel geometry to model the (2--1)/(5--4) ratios. Using the RADEX offline distribution\footnote{http://www.sron.rug.nl/$\sim$vdtak/radex/} we estimated the kinetic temperature (T$_{kin}$) and/or the volume density ($n$) from the observed line ratio. The input parameters were the background radiation field (the CMB temperature of 2.73 K) and the line width (5 km s$^{-1}$). In order to constrain the SiO column density, N(SiO), we use the observed integrated line intensity of the SiO (5--4) emission (1.5 and 4.4 K km s$^{-1}$, for B0i and B1a, respectively). Then, by running the LVG for different N(SiO), we found the best N(SiO) that better match with the SiO (5--4) brightness temperature and the (2--1)/(5--4) ratio. 

\begin{figure}
\includegraphics[angle=-90,width=9cm]{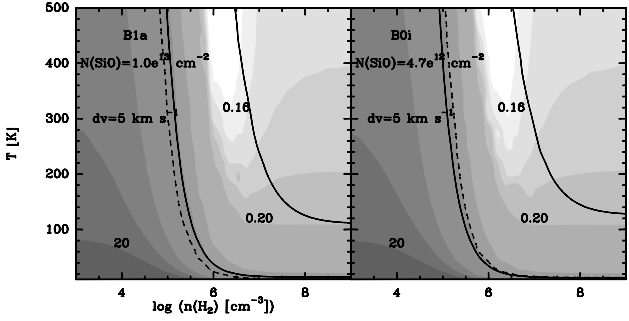}
\caption{SiO (2--1)/(5--4) ratio as a function of T$_{kin}$ and n(H$_2$) from the LVG modeling (grey scale). The panels show the case of B1a and B0i. The observed line ratios (1.7 at B1a and 1.0 at B0i) are shown as a dashed curve, while the SiO (5--4) intensities (4.4 and 1.5 K km s$^{-1}$, for B1a and B0i, respectively) are represented by the solid lines. The line ratios plotted in grey are 0.16, 0.17 0.18, 0.2, 0.5, 1, 5, 10, 20 (some of them indicated with numbers).}
\label{lvg-sio}
\end{figure}

In Fig. \ref{lvg-sio} we show the LVG results for the two cases of interest. The LVG results (Fig. \ref{lvg-sio}) shows that the (2--1)/(5--4) ratio depends on the density and less sensitive to the temperature if the density is lower than 10$^{6.5}$ cm$^{-3}$. On the other hand, the (2--1)/(5--4) ratio becomes sensitive to the temperature if the density is higher than 10$^{6.5}$ cm$^{-3}$. We found that the observed ratios yield similar solutions for both positions, with a density of $n \sim$10$^5$ to 10$^6$ cm$^{-3}$. Although the uncertainties on the line intensity and ratio are larger at B0i, these uncertainties do not affect the solution significantly (smaller than a factor of three). It should be noted that the N(SiO) at B1a is twice as higher than that at B0i. The density obtained here is comparable with 3 $\times$ 10$^5$ cm$^{-3}$ derived by \citet{Nisini07} as an averaged density including the low-velocity component. Nisini et al. also found that the physical parameters of the high-velocity component are different from those averaged over all the emitting gas; the (8--7)/(5--4) ratio at V$_{\rm LSR}$ $\sim$ -15 km s$^{-1}$ required either a higher density of $\sim$5 $\times$ 10$^6$ cm$^{-3}$ or a higher temperature of T$_K$ $>$ 500 K. Although our results do not support the high density of $>10^{6.5}$ cm$^{-3}$, they do not exclude the high temperature solution. The kinetic temperatures derived from other warm gas tracers such as high$-J$ transitions of CO, H$_2$, and H$_2$O also support the presence of warm gas component with $\sim$ 500 K (e.g. Nisini et al. 2007, 2010). We note that the density and temperature of the high-velocity clump are similar to those of the EHV bullets in the highly-collimated outflows such as HH211 and L1448C, n $\sim$ 10$^5$--10$^6$ cm $^{-3}$ and T$_K$ $\geq$ 300 K \citep[]{Nisini02,Nisini07,Hirano06,Palau06}. This implies that the high velocity emission in the B1a and B0i clumps has a common origin as that in the EHV bullets.

\begin{figure}
\centering
\includegraphics[angle=-90,width=14cm]{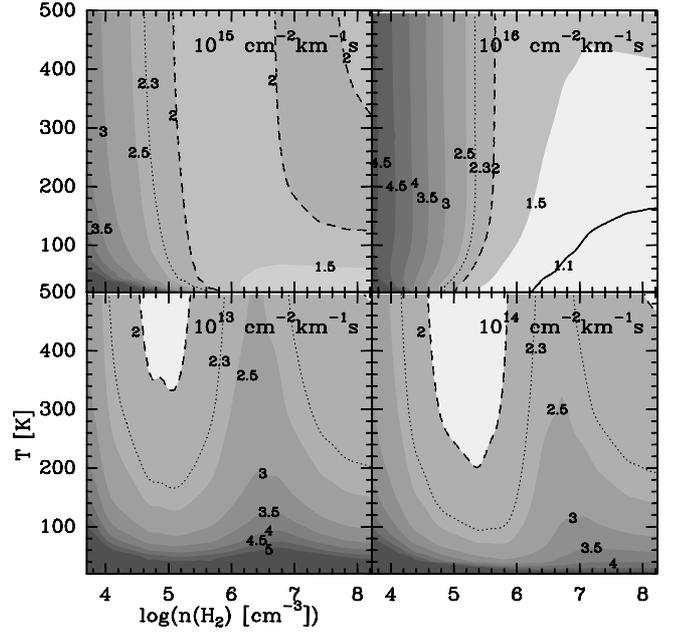}
\caption{LVG for the H$_2$CO emission in L1157. The H$_2$CO 3(0,3)--2(0,2)/ 3(2,2)--2(2,1) observed ratios are: 1.1 (solid), 2.0 (dashed), and 2.3 (dotted). Each panel shows the N(H$_2$CO)/$\Delta$v parameter used for the LVG calculation. The grey scale with numbers indicates the line ratios calculated from the LVG model.
\label{lvg-h2co}}
\end{figure}

\subsubsection{H$_2$CO emission}

The H$_2$CO 3(0,3)--2(0,2)/3(2,2)--2(2,1) ratio is sensitive to temperature in the range of 50--200 K \citep[]{vandis93,Mangum93}. These two transitions were detected in the clumps B0h, B0d, and B1a. The line ratios, 3(0,3)--2(0,2)/3(2,2)--2(2,1), averaged over the 5$''$ beam, measured at the center of these clumps, are 2.3$\pm$0.8 in B0h, 2.0$\pm$0.4 in B0d, and 1.1$\pm$0.4 in B1a. It is likely that both transitions should have significant missing flux. Since there is no available single-dish data for these transitions, it is difficult to estimate how much flux was missed. However, since the two line transitions were observed simultaneously with the same uv sampling, the line ratio can be reliable for spatially compact components. In general, the missing flux in the lower excitation line is expected to be larger than that in the higher excitation line, because the lower excitation line is more extended than the higher excitation line. Therefore, the line ratio derived here can be the lower limit, if the source is extended. Despite these limitations that would prevent a proper LVG analysis of the H$_2$CO lines, in the next we present the general trend of such analysis with the information available to us.

We have run LVG models to reproduce the H$_2$CO 3(0,3)--2(0,2)/3(2,2)--2(2,1) line ratios observed at these three clumps. In order to show the general trend of the results of the LVG for the H$_2$CO case, Fig. \ref{lvg-h2co} presents the temperature versus density plot for different N(H$_2$CO)/$\Delta$v parameters. As seen from the figure, the ratio is sensitive to temperature for N(H$_2$CO)/$\Delta$v between 10$^{13}$ and 10$^{14}$ cm$^{-1}$ km$^{-1}$ s. On the other hand, the line ratio is sensitive to densities for N(H$_2$CO)/$\Delta$v between 10$^{15}$ and 10$^{16}$ cm$^{-1}$ km$^{-1}$ s. Assuming a value of the order of 10$^{14}$ cm$^{-2}$ for N(H$_2$CO) (e.g. Bachiller \& Perez Gutierrez 1997) and a linewidth of 4 km s$^{-1}$ (the linewidth measured at these positions), the N(H$_2$CO)/$\Delta$v becomes 2.5$\times$10$^{13}$ cm$^{-2}$ km$^{-1}$ s (i.e. bottom-left panel in Fig. \ref{lvg-h2co}). In this case, the H$_2$CO line ratios suggest that the gas temperature in B0d and B0h are $>$ 350 K and $>$ 160K, respectively . On the other hand, the line ratio at B1a position has only solution in the extreme model with N(H$_2$CO)/$\Delta$v of 10$^{16}$ cm$^{-2}$ km$^{-1}$ s (top-right panel of Fig. \ref{lvg-h2co}), which then turns into a value of N(H$_2$CO) that is two orders of magnitude higher than the values derived from the single-dish observations. Such a large difference in the column density could be explained if the filling factor of the single-dish observations was very low (i.e. the emitting region is much smaller as compared to the $\sim$20$"$ beam). In order to explain the difference of two orders of magnitude, the size of the emitting region should be $\sim$2$"$ (in diameter). On the other hand, the observed low line ratio could be explained if the missing flux is more significant in the 3(0,3)--2(0,2) transition. It is natural that the 3(0,3)--2(0,2) transition in the lower energy level is more spatially extended than the 3(2,2)--2(2,1) transition. In such a case, most of the 3(0,3)--2(0,2) emission is resolved out, resulting in the low flux of 3(3,2)--2(0,2) line and very low line ratio.


\begin{figure}
\includegraphics[angle=-90,bb=44 153 562 678,width=9cm]{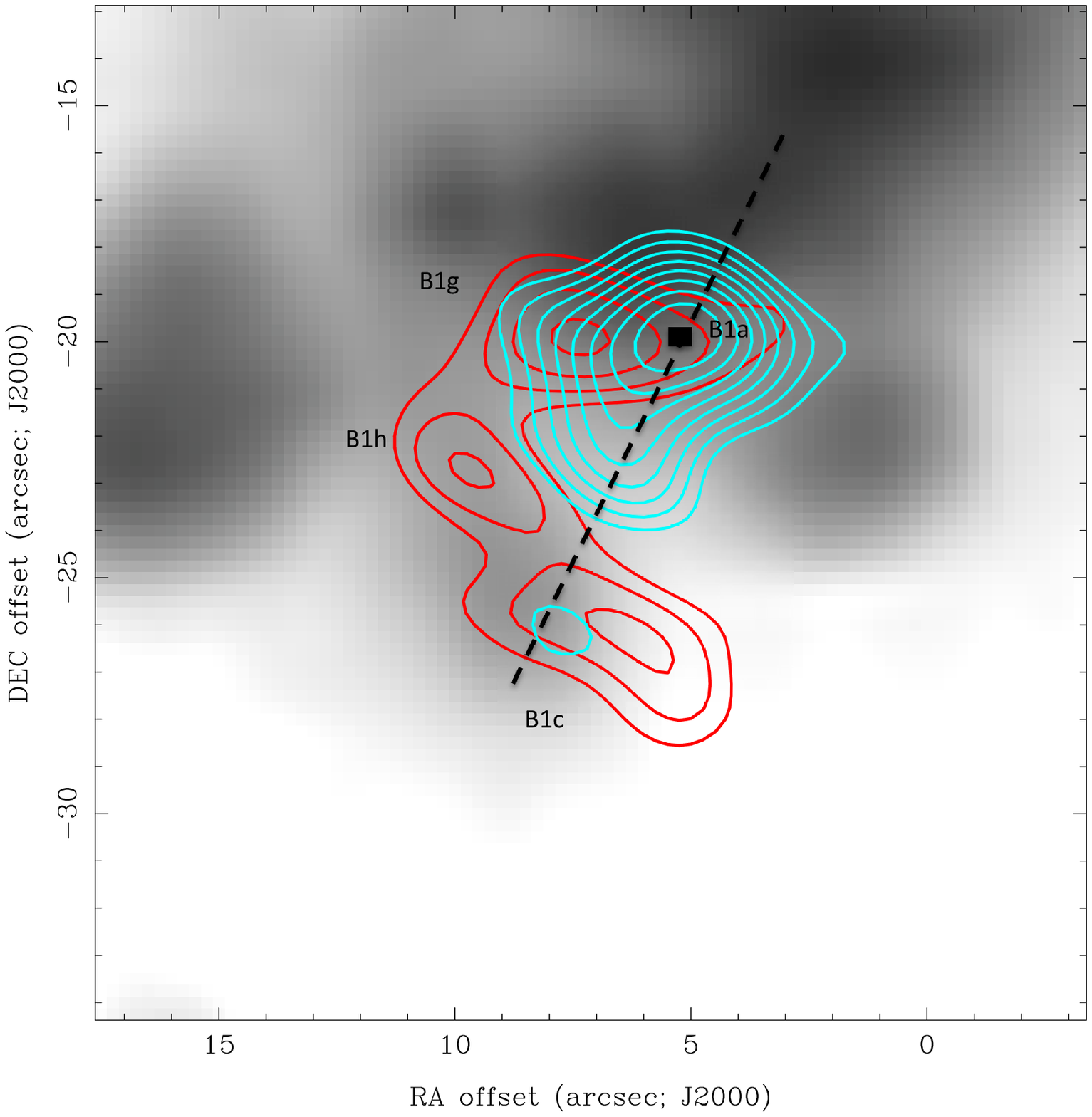}
\caption{Low velocity (red contours) and high velocity (cyan contours) SiO(5--4) emission at the B1 position, overlaid on the 4.5$\mu$m emission from IRAC/Spitzer. Dashed line indicates the axis (P.A. of 161 degrees) of the inner cavity identified in the CO (1--0) by Gueth et al. (1996). The PV diagram in Figure \ref{pv-plot} was made along this cut. Black square shows the reference position (i.e. offset=0) of the PV diagram in Figure \ref{pv-plot}.
\label{B1-struct}}
\end{figure}

\subsection{The SiO emission at B1}

\begin{figure}
\includegraphics[angle=0,width=8cm]{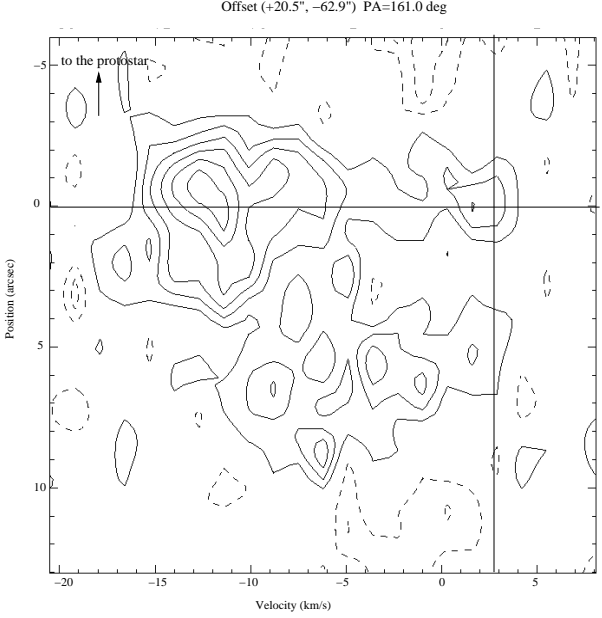}
\caption{Position-velocity diagram of the SiO(5--4) emission at the B1 position. Offset is measured from the center of the B1a clump. The cut is along the axis of the inner cavity (dashed line in Fig. 13). Vertical line indicates the cloud velocity (V$_{\rm LSR}$= 2.7 km s$^{-1}$), while horizontal line the reference position (offset=0).
\label{pv-plot}}
\end{figure}

Figure \ref{B1-struct} displays a close-up view of the high- and low-velocity SiO emission at the B1 position, taken from Fig.\ref{two-range-sio}. It is seen that the high-velocity clump B1a is surrounded by three low-velocity clumps, with two of them (B1c and B1h) located downstream of B1a. Indeed the clumpy structure of the low-velocity SiO emission at the B1 position is a remarkable finding of the present observations. The position-velocity (PV) diagram along the axis of the inner cavity (P.A. 161 deg.) also exhibits the higher velocity emission at the position closest to the protostar and the lower velocity in the downstream (Figure \ref{pv-plot}). Such a velocity structure is different from that expected in a single bow shock, in which the highest velocity appears at the tip followed by a low-velocity "wake" or "bow wing" \citep[e.g.,][]{Lee00}. If the jet axis is close to the plane of the sly (as is the case for the L1157 outflow), the velocity dispersion produced by a single bow shock decreases significantly in the post shocked region closer to the protostar \citep[see Fig. 20 of][]{Lee00}. However, the observed PV diagram exhibits the opposite trend, with the largest dispersion at the position closer to the protostar. The PV plot implies that the high-velocity clump B1a is decelerated at the interface with the low-velocity clump B1c. As discussed previously, the low-velocity clumps are likely to be the part of the spatially extended structure, which is shocked ambient gas. It is likely that the momentum transfer from high-velocity component to the ambient gas occurs at their interface.

The line profile of the high-velocity component observed with the SMA shows the maximum intensity close to the highest blue-shifted velocity with a gradual wing toward the lower velocity (Figure \ref{sio-spec-B1}). This line profile is similar to those produced by the C-type shock models calculated by \citet[]{Schilke97}. The line profile also resemble the model calculations by \citet{isaskun09} for a single shock with an age of few hundred years. This kind of line profiles are observed in the extremely high velocity (EHV) bullets of jet sources such as L1448C and HH211 \citep[e.g.,][]{Nisini07}. The similarity with the line profiles in EHV jets may suggest that the B1a clump is the possible counterpart of the EHV bullets seen in outflows like L1448C. An additional support to this suggestion is the size and mass of the high velocity clump. Since $\sim$ 80\% flux of the high-velocity SiO emission is recovered with the SMA at B1 position, the high-velocity emission is considered to come from the compact region with a size close to that of B1a clump. The size of the B1a clump as measured by a two dimensional Gaussian fit is $\sim$ 2000 $\times$ 1000 AU. Thus assuming the H$_2$ density of 10$^5$ - 10$^6$ cm$^{-3}$ (as derived from the LVG analysis), the corresponding mass is $\sim$6$\times$10$^{-5}$ - 6$\times$10$^{-4}$ M$_{\odot}$. This value is comparable to the EHV bullets in IRAS 04166$+$2706 \citep[see Table 1 of][]{Santiago09}. In summary, the line profile and physical parameters (such as density and mass) of the B1a clump suggest that this compact and high-velocity clump has a common origin as the EHV bullets observed in the jet-like outflow sources. The PV diagram at the B1 position implies that the EHV bullet is running into the dense ambient material at this position. As shown in section 4.1.1, the B0i clump also has similar physical properties as B1a. Therefore, the B0i clump is also likely to be an EHV bullet, although the missing flux at this position is uncertain.

\subsection{The SiO, H$_2$CO, and CH$_3$OH spatial distribution}


H$_2$CO and CH$_3$OH are known to be formed efficiently on grain surfaces through successive hydrogenation \citep[e.g.,][]{Tielens97,Charn97,Wata02}. Observationally, H$_2$CO and CH$_3$OH show significant abundance enhancement in shocked gas, which is likely to be originated from evaporation of grain mantles \citep[e.g.,][]{Bachiller95,Avery96,Schoi04}. As shown in Fig. \ref{two-range-h2co}, the two H$_2$CO transitions and the CH$_3$OH 4(2,2)--3(1,2)E line exhibit a similar spatial distribution, that is the walls of the U-shaped feature, suggesting that the chemistry of both molecular species are linked. However, the detailed structure within this cavity shows differences from one molecular tracer to the other. For example, if we compare the spatial distribution of the CH$_3$OH and H$_2$CO lines with similar energy levels (i.e. CH$_3$OH 4(2,2)--3(1,2) with E$_u$ $\sim$ 38 K and H$_2$CO 3(0,3)--2(0,2) with E$_u$ = 21 K), the CH$_3$OH clump (B1f) is located to the west of the apex, while the H$_2$CO clump (B1a) is located to the east of the apex. The east-west asymmetry of the CH$_3$OH was also reported in the (2$_K$-1$_K$) lines by Benedettini et al. (2007). The trend observed in the CH$_3$OH 4(2,2)--3(1,2), which is brighter in the western parts of the cavity and the apex,  is consistent with that observed in the 2$_K$-1$_K$ lines, supporting the idea of higher CH$_3$OH abundance in the western part of the cavity. On the other hand, a comparison between two transitions of H$_2$CO implies the difference in physical condition between the eastern and western parts of the cavity. As shown in section 4.1.2, the temperature of the western cavity is likely to be higher than that of the western cavity.

The H$_2$CO and CH$_3$OH lines trace well the mid-IR U-shaped structure. This U-shaped structure might be a wing of the bow shock whose tip is a bright SiO clump, B1a. On the other hand, the U-shaped structure could be formed by means of multiple shocks. As shown Fig. 5 of \citet{Takami11}, the northern lobe of the L1157 outflow contains a number of mid-IR knots along the S-shaped emission ridge. This implies that the multiple ejection events, the directions of which have varied by means of precession, have contributed to form the lobe. On the other hand, only two bright knots are identified in the southern lobe. If the ejection events themselves were symmetric, the counterparts of the 5--6 knots in the northern lobe (A2 and A3 groups in Takami et al. 2011) might have contributed to form the U-shaped cavity in the southern lobe. It is likely that the B0d clump, which is bright in H$_2$CO and CH$_3$OH lines originates from the interaction between the ejecta and ambient material. The lack of SiO in B0d clump is probably due to the low shock velocity, which is also supported by the low velocity of the H$_2$CO and CH$_3$OH emission.

The abundance of SiO is also known to be enhanced in shocked gas. The gas-phase SiO is considered to originate from silicon-bearing species that are sputtered from grains (cores or mantles) and injected into the gas phase in the form of neutral Si, SiO, SiO$_2$, or SiH$_4$ \citep[]{Schilke97,Gusdorf08a}. In Fig. \ref{3moles} we show the total integrated emission of the SiO (5--4) and H$_2$CO 3(0,3)--2(0,2) transitions, convolved to the same angular resolution of 5\arcsec. It is obvious that the spatial distribution of the SiO is different from that of the H$_2$CO (and therefore to that of CH$_3$OH, since its distribution is similar to H$_2$CO). The most prominent SiO clumps are located  at the apex of the U-shaped structure, and along the faint mid-IR arc-like feature outside of the U-shaped structure. In addition, these major SiO features appear in the high velocity range in which no H$_2$CO nor CH$_3$OH emission has been detected. {If the arc-like feature seen in the mid-IR corresponds to the wall of an outer cavity, the B0f and B0i clumps might be interacting with the wall of the outer cavity.

Fig. 9 also shows that each of the three SiO clumps is associated with a H$_2$CO clump, H$_2$CO being systematically closer to the protostar. In addition, at the positions of the H$_2$CO clumps, the low velocity SiO is also detected {(although B0h is only 3$\sigma$ in the SiO)}. These results imply the possibility that the H$_2$CO and the low velocity SiO trace the wakes of the shocks whose tips are the high-velocity SiO clumps. It is likely for the H$_2$CO in the B1a clump. However, it is unlikely to be in the cases of the B0g and B0h clumps, because the spatial distribution of the H$_2$CO 3(0,3)--2(0,2) emission is elongated along the eastern wall of the U-shaped cavity, and because B0g and B0h are not located on the lines between high-velocity SiO clumps, B0f and B0i, respectively, and the protostar. Therefore, it is more natural to consider that the H$_2$CO emission at B0g and B0h delineate the wall of the U-shaped cavity as in the case of the CH$_3$OH (2$_K$-1$_K$) emission reported by Benedettini et al. (2007).

\section{Summary and Conclusions}

The principal results of the SMA observations at 1.4 mm towards the blue-lobe of the L1157 outflow are summarized as follows:

\begin{itemize}

\item Four molecular transitions, SiO (5-4), H$_2$CO 3(0,3)--2(0,2), H$_2$CO 3(2,2)--2(2,1), and CH$_3$OH 4(2,2)--3(1,2)E were detected at B1 and B0 positions in the blue lobe. None of these lines shows significant emission at the protostellar position and the B2 position.

\item The H$_2$CO and CH$_3$OH lines trace (or delineate) the U-shaped structure seen in the mid-IR. The overall distribution of both species are similar.

\item The spatial distribution of the SiO 5-4 is significantly different from those of H$_2$CO and CH$_3$OH. The SiO (5-4) emission is brightest at the B1 position, which corresponds to the apex of the U-shaped structure. There are two compact SiO clumps along the mid-IR arc-like feature to the east of the B0 position.


\item At the B1 position, the SMA recovered ~80\% of the single-dish flux in the high-velocity range, suggesting that the high velocity emission is confined in a compact clump of $\sim$ 2000$\times$1000 AU size. The gas density of this clump derived from the SiO (2--1)/(5--4) line ratio is $\sim$10$^5$ -- 10$^6$ cm$^{-3}$, which is comparable to those of the EHV bullets. The line profile at this position is also similar to those seen in the EHV bullets; the line peaks at the highest blue-shifted velocity with a gradual wing toward the lower velocity. It is likely that the B1a clump has similar properties as the EHV bullets.

\item On the other hand, the SMA recovered only $\sim$10\% of the flux in the low-velocity range. This indicates that the low-velocity SiO is spatially extended.

\item The velocity structure around the B1 position is different from that expected in a single bow shock. It is likely that the high-velocity clump B1a is an EHV bullet running into the dense ambient material and transferring its momentum. 


\end{itemize}

\begin{acknowledgements}
 
We wish to thank all the SMA staff in Hawaii, Cambridge, and Taipei for their enthusiastic help during these observations. We also thank Mario Tafalla for providing us with the single-dish SiO(5--4) spectra and for his useful comments. AGR acknowledge ASIAA, UNAM, MPIfR, and INAF/ASI for their support during this research project. N. Hirano is supported by NSC grant 99-2112-M-001-009-My3. S.-Y. Liu is supported by NSC grant 99-2112-M-001-025-MY2.

\end{acknowledgements}


\bibliography{biblio}


\end{document}